\documentclass[12pt]{article}

\def\be{\begin{equation}}
\def\ee{\end{equation}}

\usepackage{euscript,latexsym}

\begin{document}

\title{{\bf An Attack to Quantum Cryptography from Space}}
\author{Igor V. Volovich\thanks{Permanent address: Steklov Mathematical
Institute, Gubkin St.8, GSP-1, 117966, Moscow, Russia;
volovich@mi.ras.ru}\\~\\ {\it Department of Mathematics,
Statistics and Computer Sciences}\\ {\it University of Vaxjo,
S-35195, Sweden}\\  }
\date{}
\maketitle

\begin{abstract}

The promise of secure cryptographic quantum key distribution
schemes is based on the use of quantum effects in the spin space.
We point out that in fact in many current quantum cryptography
protocols the space part of the wave function is neglected.
However exactly the space part of the wave function describes the
behaviour of particles in ordinary real three-dimensional  space.
As a result such schemes can be secure against eavesdropping
attacks in the abstract spin space but  could be insecure in the
real three-dimensional space. We discuss an approach to the
security of quantum key distribution in space by using Bell's
inequality and a special preparation of the space part of the wave
function.
\end{abstract}

\newpage
\section{Introduction}
It is now generally accepted that techniques of quantum
cryptography can allow secure communications between distant
parties  \cite{BB} - \cite{ZLG}. The promise of secure
cryptographic quantum key distribution schemes is based on the use
of  quantum entanglement in the spin space and  on quantum
no-cloning theorem. An important contribution of quantum
cryptography is a mechanism for detecting eavesdropping.

In the present note we point out that in fact in many current
quantum cryptography protocols the space part of the wave function
is neglected. However exactly the space part of the wave function
describes the behaviour of particles in ordinary real
three-dimensional
 space. As a result such schemes can be secure against
eavesdropping attacks in the abstract spin space but could  be
insecure in the real three-dimensional space. We discuss an
approach to the security of quantum key distribution in the real
space by using Bell's inequality and a special preparation of the
space part of the wave function.

Bell's theorem \cite{Bel} states that there are quantum
correlation functions that can not be represented as classical
correlation functions of separated  random variables. It has been
interpreted as incompatibility of the requirement of locality with
the statistical predictions of quantum mechanics \cite{Bel}.
Bell's theorem constitutes an important part in quantum
cryptography protocols \cite{Eke}.

It was mentioned in  \cite{Vol} that the space dependence of the
wave function is neglected in discussions of the problem of
locality in relation to Bell's inequalities. However it is the
space part of the wave function which is relevant to the
consideration of the problem of locality. Similar remark one can
apply to quantum teleportation  \cite{Vol2}. It was shown
 \cite{Vol} that the
space part of the wave function leads to an extra factor in
quantum correlation and as a result the ordinary proof of Bell's
theorem  fails in this case.

It follows that   proofs of the security of quantum cryptography
schemes which neglect the space part of the wave function could
fail against attacks in the real three-dimensional space. We will
discuss how one can try to improve the security of quantum
cryptography schemes in space  by using  a special preparation of
the space part of the wave function.

\section {Quantum Key Distribution}
Ekert \cite{Eke} showed that one can use the EPR correlations to
establish a secret random key between two parties ("Alice" and
"Bob"). Bell's inequalities are used to check the presence of an
intermediate eavesdropper ("Eve"). There are two stages to the
Ekert protocol, the first stage over a quantum channel, the second
over a public channel.

 The quantum channel consists of a source
that emits pairs of spin one-half particles, in a singlet state.
The particles fly apart towards Alice and Bob, who, after the
particles have separated, perform measurements on spin components
along one of three directions, given by unit vectors $a$ and $b$.
In the second stage Alice and Bob communicate over a public
channel.They announce in public the orientation of the detectors
they have chosen for particular measurements. Then they divide the
measurement results into two separate groups: a first group for
which they used different orientation of  the detectors, and a
second group for which they used the same orientation of the
detectors. Now Alice and Bob can reveal publicly the results they
obtained but within the first group of measurements only. This
allows them, by using Bell's inequality, to establish the presence
of an eavesdropper (Eve). The results of the second group of
measurements can be converted into a secret key.

\section {Bell's Inequality}
In the  presentation of Bell's theorem we will follow  \cite{Vol}
where one can find also more references. Consider a pair of spin
one-half particles formed in the singlet spin state and moving
freely towards Alice and Bob. If one neglects the space part of
the wave function  then the quantum mechanical correlation of two
spins in the singlet state $\psi_{spin}$ is
\be
\label{eqn1}
 E_{spin}(a,b)=\left<\psi_{spin}|\sigma\cdot a \otimes\sigma\cdot
b|\psi_{spin}\right>=-a\cdot b
 \ee
 Here $a$ and $b$ are two
unit vectors in three-dimensional space and
$\sigma=(\sigma_1,\sigma_2,\sigma_3)$ are the Pauli matrices.
Bell's theorem states that the function $ E_{spin}(a,b)$
(\ref{eqn1})
 can not be represented in the
form
\be
\label{eqn2} P(a,b)=\int \xi (a,\lambda) \eta (b,\lambda)
d\rho(\lambda) \ee
 Here $ \xi (a,\lambda)$ and $  \eta(b,\lambda)$ are random fields on the sphere,
 $| \xi (a,\lambda)|\leq 1$, $ | \eta (b,\lambda)|\leq 1$ and
 $d\rho(\lambda)$ is a positive probability measure, $ \int d\rho(\lambda)=1$.
The parameters $\lambda$ are interpreted as hidden variables in a
realist theory. One supposes that Eve is described by these
variables.

One has the following  Bell-Clauser-Horn-Shimony-Holt (CHSH)
inequality
\be
\label{eqn3}
 |P(a,b)-P(a,b')+P(a',b)+P(a',b')|\leq 2
\ee
 From the other hand there are such vectors
$(ab=a'b=a'b'=-ab'=\sqrt{2}/2)$ for which one has
\be
\label{eqn4}
 | E_{spin}(a,b)- E_{spin}(a,b')+ E_{spin}(a',b)+ E_{spin}(a',b')|=2\sqrt{2}
\ee Therefore if one supposes that $ E_{spin}(a,b)=P(a,b)$ then
one gets the contradiction.

\section {Localized Detectors}

In the previous section the space part of the wave function of the
particles was neglected. However exactly the space part is
relevant to the discussion of locality. The complete wave function
is $\psi =(\psi_{\alpha\beta}({\bf r}_1,{\bf r}_2))$
%\be
%\label{eqn5} \psi({\bf r}_1,{\bf r}_2)=\psi_{spin}\psi_{1}({\bf
%r}_1)\psi_{2}({\bf r}_2) \ee
where $\alpha$ and $\beta $ are spinor indices and ${\bf r}_1$ and
${\bf r}_2$ are vectors in three-dimensional space.

We suppose that Alice and Bob have detectors which are located
within the two localized regions ${\cal O}_A$ and ${\cal O}_B$
respectively, well separated from one another. Eve has a detector
which is located within the region ${\cal O}_E$. Quantum
correlation describing the measurements of spins by Alice and Bob
at their localized  detectors is
\be
\label{eqn6}
 E(a,{\cal O}_A,b,{\cal O}_B)=\left<\psi| \sigma\cdot a   P_{{\cal O}_A}
 \otimes  \sigma\cdot b  P_{{\cal O}_B} |\psi\right>
 \ee
Here $P_{{\cal O}}$ is the projection operator onto the region
${\cal O}$.

Let us consider the case when the wave function has the form
$\psi=\psi_{spin}\phi({\bf r}_1,{\bf r}_2)$. Then one has
\be
\label{eqn7}
 E(a,{\cal O}_A,b,{\cal O}_B)=g ({\cal O}_A,{\cal O}_B)
  E_{spin}(a,b)
 \ee
 where the function
\be
\label{eqn8}
 g ({\cal O}_A,{\cal O}_B)=\int_{{\cal O}_A \times {\cal O}_B}|\phi({\bf
 r}_1,{\bf
 r}_2)|^2 d{\bf r}_1d{\bf r}_2
 \ee
 describes correlation of particles in space.
 Note that one has
\be
\label{eqn8g} 0\leq g ({\cal O}_A,{\cal O}_B)\leq 1
 \ee

{\bf Remark.} In relativistic quantum field theory
 there is no nonzero
strictly localized projection operator that annihilates the
vacuum. It is  a consequence of the Reeh-Schlieder theorem
 \cite{Lue}. Therefore, apparently, the function
 $ g ({\cal O}_A,{\cal O}_B)$ should be always strictly
smaller than 1. I am grateful to W. Luecke for this remark.

We will interpret Eve as  a  hidden variable in a realist theory
and  will study whether
 the quantum correlation (\ref{eqn7}) can be represented in the
 form (\ref{eqn2}). More exactly one inquires whether one can write
 the representation
\be
\label{eqn9}
 g({\cal O}_A,{\cal O}_B)E_{spin}(a,b)=\int \xi (a,{\cal O}_A,\lambda)
 \eta (b,{\cal O}_B,\lambda) d\rho(\lambda)
 \ee

Now from (\ref{eqn3}), (\ref{eqn4}) and (\ref{eqn9}) one can
obtain the  following inequality
\be
\label{eqn10}
 g ({\cal O}_A,{\cal O}_B)\leq 1/\sqrt 2
 \ee
 If the inequality (\ref{eqn10}) is valid for  regions  ${\cal
O}_A$ and ${\cal O}_B$ which are well separated from one another
 then there is no violation
of the CHSH inequalities (\ref{eqn3}) and therefore Alice and Bob
can not detect the presence of an eavesdropper. From the other
side, if for a pair of well separated regions ${\cal O}_A$ and
${\cal O}_B$ one has
\be
\label{eqn11}
 g ({\cal O}_A,{\cal O}_B) > 1/\sqrt 2
  \ee
then it could be a  violation of the realist locality in these
regions for a given state. Then, in principle, one can hope to
detect an eavesdropper in these circumstances.

Note that if we set $ g({\cal O}_A,{\cal O}_B)=1$ in (\ref{eqn9})
as it was done in the original proof of Bell's theorem, then it
means we did a special preparation of the states of particles to
be completely localized inside of detectors. There exist such well
localized states (see however the previous Remark) but there exist
also another states, with the wave functions which are not very
well localized inside the detectors, and still particles in such
states are also observed in detectors. The fact that a particle is
observed inside the detector does not mean, of course, that its
wave function is strictly localized inside the detector before the
measurement.  Actually one has  to perform a thorough
investigation of the preparation and the  evolution of our
entangled states in space and time if one needs to estimate the
function $ g({\cal O}_A,{\cal O}_B)$.

\section {Conclusions}

It is pointed out  in this note that  the presence of the space
part in the wave function of two particles in the entangled state
leads to a problem in the proof of the security of quantum key
distribution. To detect the eavesdropper's presence by using
Bell's inequality we have to estimate the function  $ g({\cal
O}_A,{\cal O}_B)$. Only a special quantum key distribution
protocol has been discussed here but it seems there are similar
problems in other quantum cryptographic schemes as well.

 We don't claim in this note that it is in principle impossible to increase the
detectability of the eavesdropper. However it is not clear to the
present author how to do it without a
  thorough investigation of the process of preparation of the entangled state
and then its  evolution in space and time towards Alice and Bob.

In the previous section Eve was interpreted as an abstract hidden
variable. However one can assume that more information about Eve
is available. In particular one can assume that she is located
somewhere in space in a region ${\cal O}_E.$ It seems one has to
study a generalization of the function  $ g({\cal O}_A,{\cal
O}_B)$, which depends not only on the Alice and Bob locations
${\cal O}_A$ and $ {\cal O}_B$ but also depends on the Eve
location $ {\cal O}_E$, and try to find a strategy which leads to
an optimal value of this function.

\section{Acknowledgments}

I would like to thank A.Khrennikov  and  the Department of
Mathematics, Statistics and Computer Sciences, University of
Vaxjo, where this work was done, for the warm hospitality. I am
grateful to J.L. Cereceda,   G. Garbarino, S. Goldstein, W. Hofer,
W. Luecke, N. D. Mermin, K. Svozil,
 and P. Zanardi for
correspondence on Bell's theorem.

%%%%%%%%%%%%%%%%%%%%%%%%%%%%%%%%%%%%%%%%%%%%%%%%%%%%%%%%%

\end{document}